\pdfoutput=1

\documentclass[review]{elsarticle}

\usepackage{graphicx}
\usepackage{amsmath}
\usepackage{amssymb}
\usepackage{mathrsfs}
\usepackage{subfigure}
\def\be{\begin{equation}}
\def\ee{\end{equation}}
\def\ba{\begin{array}}
\def\ea{\end{array}}
\def\bea{\begin{eqnarray}}
\def\eea{\end{eqnarray}}
\def\no{\nonumber}
\def\({\left(}
\def\){\right)}
\def\[{\left[}
\def\]{\right]}

\usepackage{lineno}

\journal{Physics Letters A}

\begin{document}
\begin{frontmatter}

\title{Chirped Lambert W-kink solitons of the complex cubic-quintic Ginzburg-Landau equation with intrapulse Raman scattering}

\author[mymainaddress,mysecondaryaddress]{Nisha}

\author[mysecondaryaddress]{Neetu Maan}

\author[mymainaddress]{Amit Goyal\corref{mycorrespondingauthor}}
\cortext[mycorrespondingauthor]{Corresponding author}
\ead{amit.goyal@ggdsd.ac.in}

\author[mysecondaryaddress2]{Thokala Soloman Raju}
\ead{solomonr\_thokala@yahoo.com}

\author[mysecondaryaddress]{C. N. Kumar}
\ead{cnkumar@pu.ac.in}

\address[mymainaddress]{Department of Physics, GGDSD College, Chandigarh 160030, India}
\address[mysecondaryaddress]{Department of Physics, Panjab University, Chandigarh 160014, India}
\address[mysecondaryaddress2]{Indian Institute of Science Education and Research (IISER) Tirupati, Andhra Pradesh 517507, India}

\begin{abstract}
In this paper, an exact explicit solution for the complex cubic-quintic Ginzburg-Landau equation is obtained, by using Lambert W function or omega function. More pertinently, we term them as Lambert W-kink-type solitons, begotten under the influence of intrapulse Raman scattering. Parameter domains are delineated in which these optical solitons exit in the ensuing model. We report the effect of model coefficients on the amplitude of Lambert W-kink solitons, which enables us to control efficiently the pulse intensity and hence their subsequent evolution. Also, moving fronts or optical shock-type solitons are obtained as a byproduct of this model. We explicate the mechanism to control the intensity of these fronts, by fine tuning the spectral filtering or gain parameter. It is exhibited that the frequency chirp associated with these optical solitons depends on the intensity of the wave and saturates to a constant value as the retarded time approaches its asymptotic value.
\end{abstract}

\end{frontmatter}


\section{Introduction}
The complex Ginzburg-Landau equation is a canonical model
for weakly nonlinear, dissipative systems and one of the
most-studied nonlinear equations in the physics community. It can
be used to describe a vast variety of nonlinear phenomena such as
superconductivity \cite{A.M.}, Bose-Einstein condensation
\cite{E.Kengne}, superfluidity \cite{A. Berti}, strings in field
theory \cite{R.J. Rivers}, liquid crystals \cite{M. C.} and lasers
\cite{M.S. Osman,Osman2}. Exact solitary pulse solutions of complex Ginzburg-Landau equation with the cubic nonlinearity are available, but these
pulses are stable only in a region where the background is
unstable \cite{akh}. However, it is possible to create stable
solitary pulses, in a region where background is also stable, with the
inclusion of an extra term, represents delayed Raman
scattering, in the model \cite{Facao3,facao4}. The existence and stability of solitary wave solutions for cubic complex Ginzburg-Landau equation is also studied in the presence of driven term \cite{malomed,raju,amit}. Recently, lot of attention has been paid to obtain exact analytical solutions for complex systems modeled by nonautonomous partial differential equations \cite{osman01,osman02,osman03,osman04,osman05}.

\par Soliton propagation in fibers with linear and nonlinear gain and spectral filtering \cite{4M. Mats} or pulse generation in fiber lasers with additive pulse mode-locking or nonlinear polarization rotation \cite{5H.A.,6J.D.} have been studied by considering complex cubic-quintic Ginzburg-Landau equation (CQGLE) as model equation. Also Kengne and Vaillancourt have used modified Ginzburg–Landau equation that describes the pulse propagation in a lossy electrical transmission
line \cite{kengne}. The CQGLE supports a class of localized solutions such as stationary solitons, sources, sinks, moving solitons and fronts with fixed velocity \cite{saarlos,akh2,crespo}. Apart from these solutions, the CQGLE also possesses the solutions with special propagation properties: pulsating, creeping, and erupting solitons \cite{J.Soto2,2001PRE,2001PLA}. The erupting solitons are those that periodically exhibits explosive instability. These solitons were found numerically \cite{J.Soto2} and also experimentally in passively mode-locked lasers \cite{2002PRL}. The effect of higher-order terms, namely, third-order dispersion, self-steepening and intrapulse Raman scattering has been investigated on erupting solitons and it is found that the explosions of an erupting soliton can be controlled or even canceled due to the inclusion of one or more higher-order terms \cite{tian,latas,M.Facao1,M.Facao2,carvalho,latas2}. Recently, work has been done to study the transitions of stationary to pulsating solutions \cite{2018} and on the selection mechanism of soliton explosions in CQGLE under the influence of higher-order terms \cite{2019}. Fac\~{a}o et al. have studied the effect of intrapulse Raman scattering (IRS) on erupting solitons of CQGLE and numerically shown the propagation of stable traveling solitons for a specific range of IRS parameter \cite{M.Facao1,M.Facao2}. Although, CQGLE is a well studied dynamical system, the exact solutions of CQGLE with IRS term have rarely appeared in the literature. However, in Ref. \cite{anki}, the authors have presented the exact stationary front solutions for CQGLE with Raman term and generalized these solutions into moving fronts using energy and momentum balance equations for particular cases by assuming either the quintic or Raman term to be zero.

\par In this work, we consider the CQGLE in the presence of IRS and report the existence of exact localized  solutions in the form of dark and front solitons. The dark solitons are presented by a new kind of kink solution in terms of Lambert W function which we shall refer to as Lambert W-kink solitons. The Lambert W function is an implicitly elementary function, also
known as the product logarithm, has rich variety of applications in
number of areas of physics, computer science, pure and applied
mathematics and ecology, and is defined as the inverse of $f(W)=W e^W$ \cite{R.M.,S.R.}. Several
well-known problems in electrostatics and in quantum mechanics can
be solved with greater ease using the notation of Lambert W function. Biswas et al. used the notation of Lambert W function to obtain soliton solutions of modified nonlinear Schr\"odinger equation using variational principle \cite{Biswas2}. This function is also used as a step potential for which the one-dimensional stationary Schr\"odinger equation is exactly solved in terms of the confluent hypergeometric functions \cite{potential}. Recently, soliton solution in the form of Lambert W function was obtained for analytically solvable parity-breaking $\phi^6$ model and the results so obtained were compared with kink of $\phi^4$ theory \cite{Azadeh}. Apart from Lambert W-kink solitons, we have also explored moving front solitons for this model. The evolution of optical solitons can be controlled by judicious choice of model parameters. The frequency chirp is found to be directly proportional to the
intensity of the wave and saturates at some
finite value as $t\rightarrow\pm\infty$. Frequency chirp is a well-known result of the
interaction of the group velocity dispersion and the nonlinear self phase modulation.
Chirp is very useful in the process of optical pulse compression and found potential applications in optical communication systems \cite{chirp1,chirp2,chirp3}. A significant work has been done on the existence of chirped solitons in the context of nonlinear optics \cite{chirp4,chirp5,Alka,goyal}.

 \section{Model Equation}
We begin our analysis by considering the complex cubic-quintic
Ginzburg-Landau equation (CQGLE) with intrapulse Raman scattering (IRS) term
    \be\label{eq1} i U_z+\frac{1}{2} U_{tt}+\gamma\vert U \vert^2 U=i\delta U+i \beta U_{tt}+i \epsilon\vert U\vert^2U-\nu\vert U \vert^4 U+i \mu \vert U \vert^4 U+T_r{(\vert U \vert^2)}_t U,\ee
where $U$ is the normalized envelope of
the pulse, $z$ and $t$ are the normalized propagation distance
and retarded time, respectively. For laser system \cite{6J.D.}, the physical meaning of various coefficients is the following:
$\delta$ is a constant gain (or loss if negative), $\beta$ describes
spectral filtering or gain dispersion, $\epsilon$ represents
nonlinear gain (or two-photon absorption if negative), $\mu$ represents a
higher order correction to the nonlinear amplification or absorption, $\nu$ is a higher order correction term to the nonlinear refractive index, $T_r$ represents the IRS coefficient and $\gamma$ represents positive Kerr effect (or negative Kerr effect if negative) .

\section{Chirped soliton-like solutions}
In order to find the exact solution of Eq. (\ref{eq1}), we choose the following ansatz
 \be\label{eq2} U(z,t)=\rho(\xi)e^{i(\phi(\xi) -k z)}, \ee
where $\xi=t-u z$ is the traveling coordinate, $\rho$ and
$\phi$ are real functions of $\xi$. Here $u = \frac{1}{v}$, where $v$ indicates the group
velocity of the pulse envelope. The corresponding intensity of the propagating pulse is given by $|U(z,t)|^2=|\rho(\xi)|^2$. The spectral changes introduced across the pulse at any distance $z$ are a direct consequence of time dependence of nonlinear phase
shift. The frequency change across the pulse is the time
derivative of phase and is given by $\delta\omega(z,t)=-\frac{\partial}{\partial t}[\phi(\xi)-k
z]=-\phi'(\xi)$. This time dependence of $\delta\omega$ is referred to as frequency
chirping. Now, substituting Eq. (\ref{eq2}) into
Eq. (\ref{eq1}), and separating out the real and imaginary parts of the equation, we obtain
the following coupled equations in $\rho$ and $\phi$,
    \be\label{eq3}u\rho\phi'+k\rho+\frac{1}{2}\rho''-\frac{1}{2}\rho\phi'^2+\gamma\rho^3=-2\beta\rho'\phi'-\beta\rho\phi''-\nu\rho^5+2T_r\rho^2\rho',\ee
    \be\label{eq4}-u\rho'+\rho'\phi'+\frac{1}{2}\rho\phi''=\delta\rho+\beta\rho''-\beta\rho\phi'^2+\epsilon\rho^3+\mu\rho^5.\ee
Assuming that the qualitative features of frequency chirp depend considerably on the exact pulse shape through the relation $\delta\omega(z,t)=-\phi'(\xi)=-(A\rho^2+B)$, where A and B are the nonlinear and constant chirp parameters, respectively,
the coupled equations given by Eq. (3) and Eq. (4) reduce to

    \be\label{eq6} \rho''+4(2\beta A-T_r)\rho^2 \rho'+4\beta B\rho'+(2uB-B^2+2k)\rho+2(u A+\gamma-A B)\rho^3+(2\nu-A^2)\rho^5=0,\ee
    \be\label{eq7} \rho''-\frac{2A}{\beta}\rho^2 \rho'+\frac{(u-B)}{\beta} \rho'+\(\frac{\delta}{\beta}-B^2\)\rho+\(\frac{\epsilon}{\beta}-2AB\)\rho^3+\(\frac{\mu}{\beta}-A^2\)\rho^5=0,\ee
for $\beta \neq 0$. By assuming the following identifications:

\be\label{eq8} M\equiv\(8\beta A-4T_r=-\frac{2A}{\beta}\),\ee

\be\label{eq9} N\equiv \(4\beta B=\frac{u-B}{\beta}\),\ee

\be\label{eq10} Q\equiv \(2uB+2k-B^2=\frac{\delta}{\beta}-B^2\),\ee

\be\label{eq11} R\equiv \(2Au-2AB+2\gamma=\frac{\epsilon}{\beta}-2AB\),\ee

\be\label{eq12} S\equiv \(2\nu-A^2=\frac{\mu}{\beta}-A^2\),\ee

Eqs. (\ref{eq6}) and (\ref{eq7}) can be mapped into a single equation
    \be\label{eq13}\rho''+M\rho^2\rho'+N\rho'+Q\rho+R\rho^3+S\rho^5=0.\ee
Solving Eqs. (\ref{eq8})-(\ref{eq12}), we obtain the constraint conditions as
     \be\label{eq14}\no A=\frac{2\beta  T_r}{1+4\beta^2},~~u=\frac{1}{A}\left(\frac{\epsilon }{2\beta }-\gamma \right),~~B=\frac{u}{1+4\beta^2}\ee
     \be\label{eq15}k=\frac{\delta }{2\beta }-u B,~~\mu =2\beta \nu.\ee
Eq. (\ref{eq13}) can be solved to obtain exact localized solution for compatible form of first-order differential equation for the function $\rho(\xi)$. In this work, we have explored the Lambert W-kink and moving front soliton solutions for this equation.
\subsection{Lambert W-kink solitons}
In order to explore exact analytical solution of Eq. (12), use shall be made of the differential equation
    \be\label{eq17}\rho'=(a^2-\rho^2)~(a-\rho),\ee
 ($a$ is a real parameter here), that admits Lambert W-kink solution of the form \cite{R.M.,Azadeh}
    \be\label{eq21}\rho(\xi)=a\(1-\frac{2}{1+W(e^{4a^2\xi+1})}\),\ee
where $W$ represents Lambert W function. The corresponding second-order differential equation for $\rho(\xi)$ reads
     \be\label{eq18}\rho''-3\rho^5+5a\rho^4+2a^2\rho^3-6a^3\rho^2+a^4\rho+a^5=0. \ee
For $\rho'$ given by Eq. (\ref{eq17}), Eq. (\ref{eq13}) is consistent with the Eq. (\ref{eq18}) by the identification of various unknown parameters as $M=-5,N=a^2,Q=2a^4,R=-4a^2,S=2$. Solving these conditions along with constraints given by Eq. (\ref{eq15}), the model coefficients and solution parameter $`a$' fixed as
    \be\label{eq19}\no T_r=\frac{5}{4}\left(1+4\beta ^2\right),~~\epsilon =\frac{11\beta \gamma} {2\left(4+5\beta^2\right)},~~\nu=1+\frac{25}{8}\beta^2,\ee
    \be\label{eq20} \delta=\frac{\gamma^2\left(1+32\beta^2\right)}{4\beta\left(4+5\beta^2\right)^2},~~a=\sqrt{\frac{-2\gamma}{4+5\beta ^2}}.\ee

\begin{figure}[ht!]
    \begin{center}
         \includegraphics[angle=0,scale=.3]{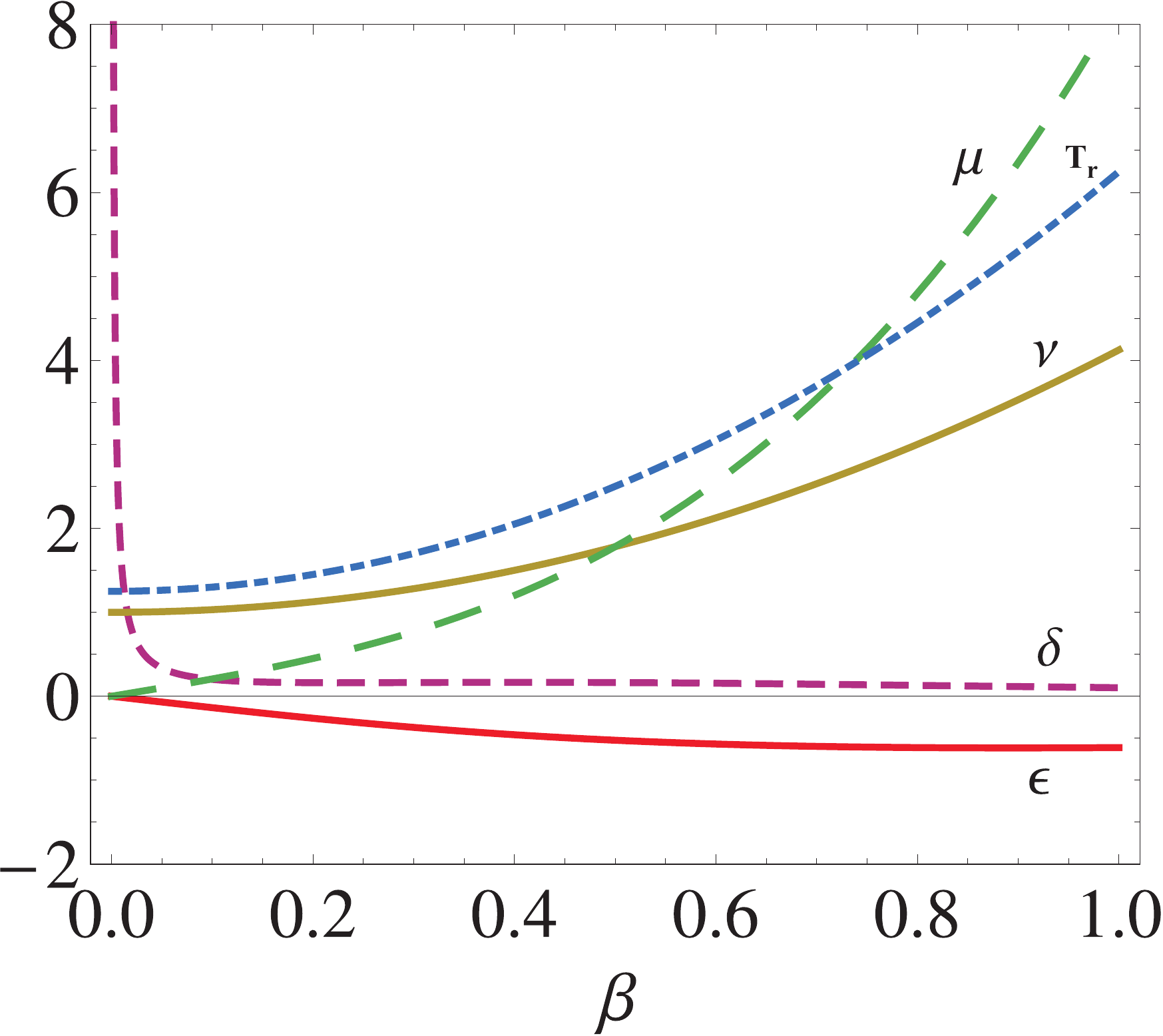}
    \caption{\label{fig01} Curves of model coefficients versus spectral filtering term $`\beta$' for $\gamma=-1$.}
    \end{center}
\end{figure}
\begin{figure}[ht!]
\begin{center}
\includegraphics[angle=0,scale=.60]{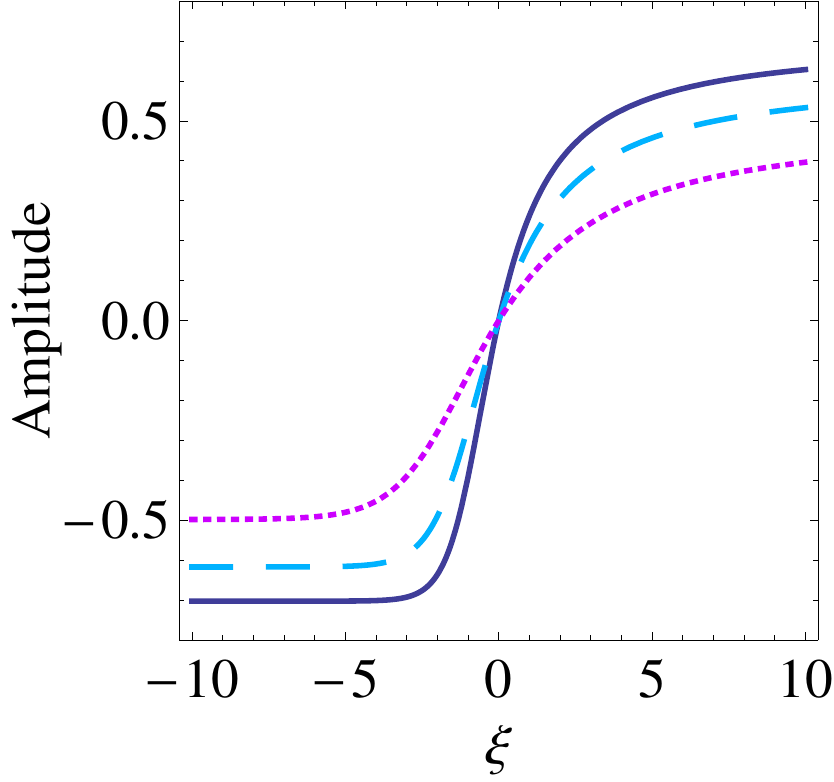}
\caption{\label{fig1} Amplitude profiles of Lambert
W-kink solution for different values of $\beta$, $\beta=0.1$ (thick line), $\beta=0.5$
(dashed line) and $\beta=0.9$ (dotted line).}
\end{center}
\end{figure}

It should be noted that $\gamma$ will take only negative values, as solution parameter $`a$' should be real, and $\beta$ can be chosen arbitrarily $(\beta \neq 0)$ while the other model coefficients depend on $\beta$ and $\gamma$. In Fig. \ref{fig01}, we have presented the allowed values of the model coefficients with $\beta$ lying in the interval $[0,1]$ for $\gamma=-1$. In Fig. \ref{fig1}, we have shown the amplitude profile of Lambert W-kink solution, given by Eq. (\ref{eq21}), for different values of the spectral filtering term $`\beta$' and $\gamma=-1$. From this plot, one can observe that kink wave has large amplitude and becomes more steep for small values of $\beta$ as solution parameter $`a$' is inversely proportional to $\beta$. It should be noted that, from Fig. \ref{fig01}, these Lambert W-kink solutions are possible only for $\delta>0, ~\epsilon<0$ and $\gamma<0$.

\begin{figure}[ht!]
    \begin{center}
        \subfigure[]{
    \includegraphics[angle=0,scale=.5]{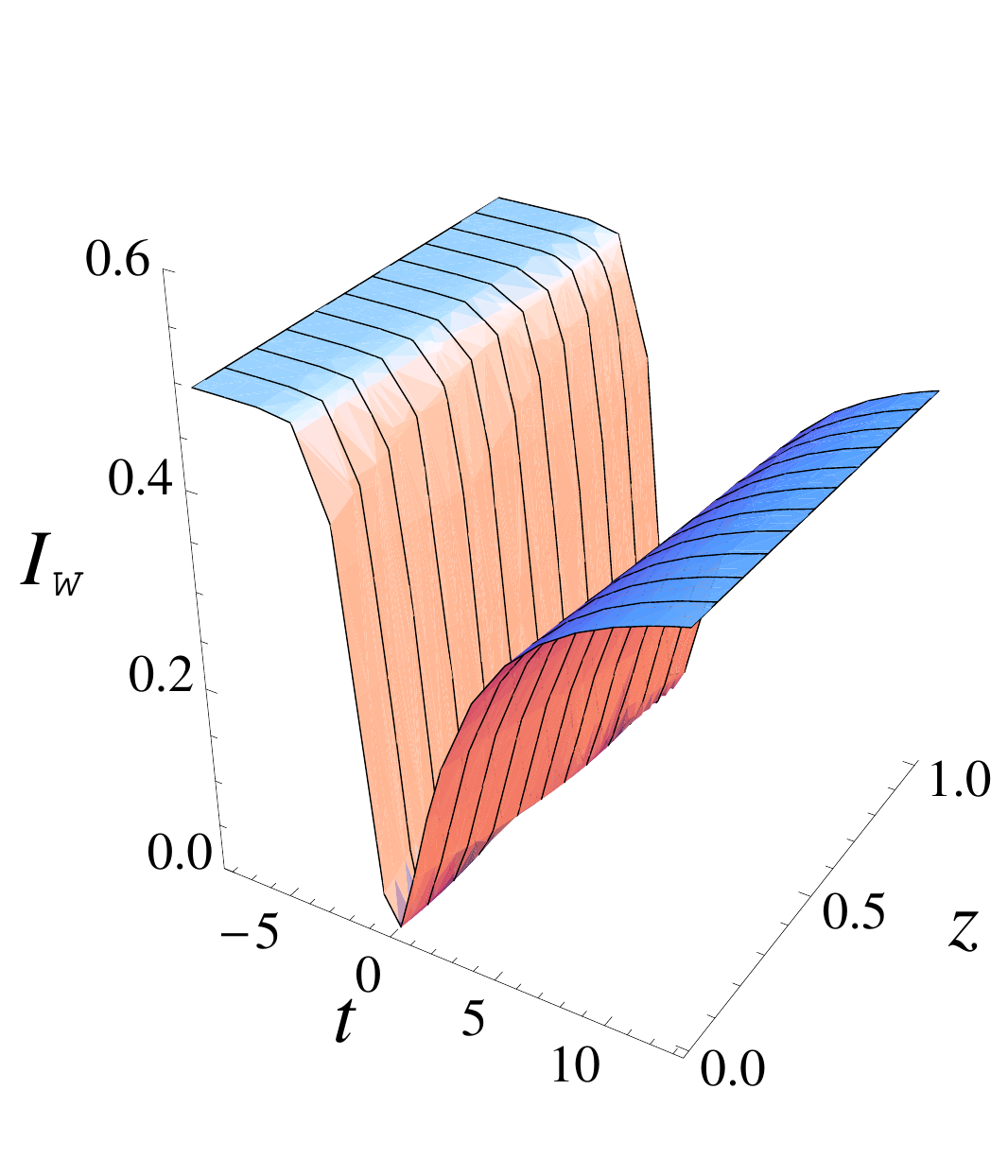}
    }
     \subfigure[]{
    \includegraphics[angle=0,scale=.5]{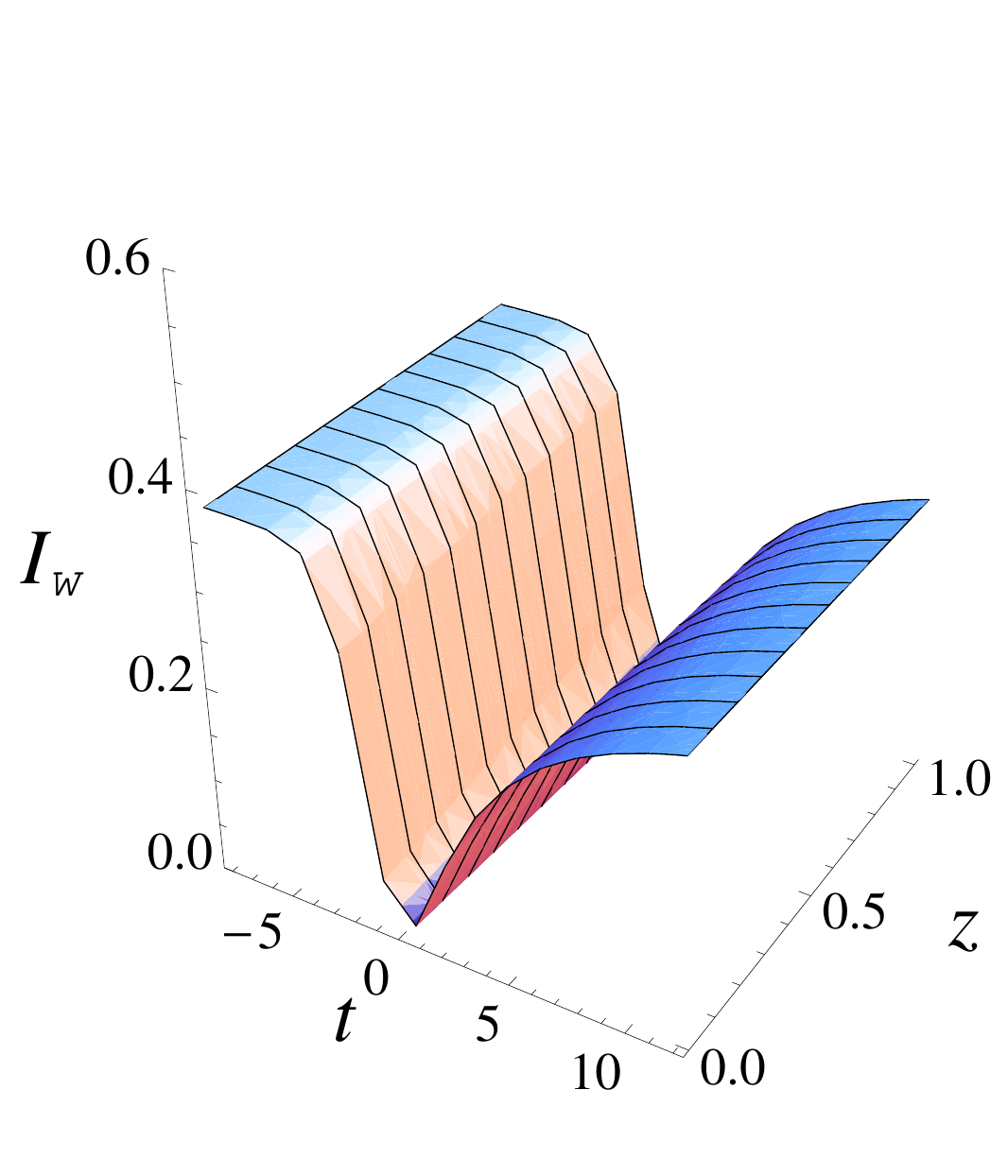}
     }
     \subfigure[]{
    \includegraphics[angle=0,scale=.7]{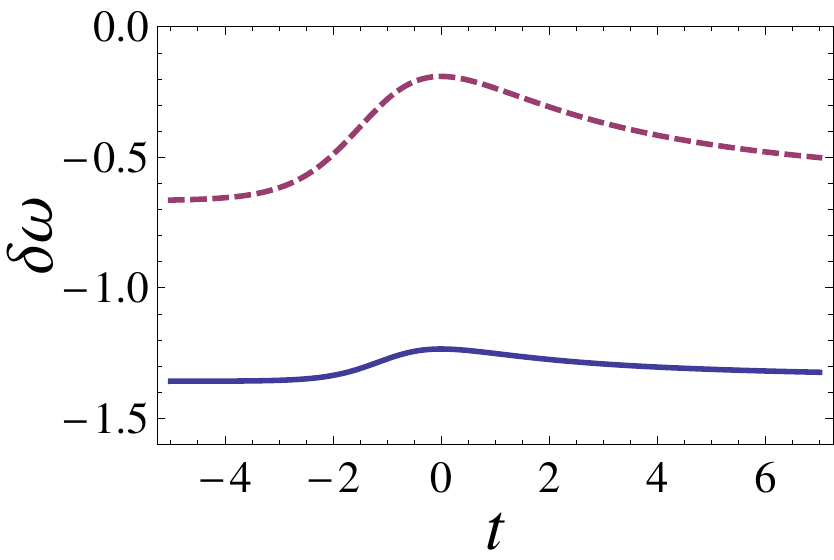}
    }
    \caption{\label{fig2}(a,b) Intensity profile of Lambert W-kink soliton for $\gamma=-1$ and different values of coefficient $\beta$, $0.1$ and $0.5$, respectively. (c) The corresponding chirp profiles
     for $\beta=0.1$ (thick line) and $\beta=0.5$ (dashed line).}
    \end{center}
\end{figure}

Using Eq. (\ref{eq21}) into the Eq. (\ref{eq2}), the intensity expression of Lambert W-kink solitons, for the model equation Eq. (\ref{eq1}), reads
    \be\label{i1}I_W(z,t)=a^2\(1-\frac{2}{1+W(e^{4a^2\xi+1})}\)^2,\ee
with chirping given by
    \be\label{i2}\delta\omega(z,t)=-\[\frac{5a^2\beta}{2}\(1-\frac{2}{1+W(e^{4a^2\xi+1})}\)^2-\frac{\gamma }{2\beta \left(4+5 \beta ^2\right)}\].\ee
The intensity profile of Lambert W-kink soliton is depicted in the Fig. \ref{fig2}(a,b) for $\gamma=-1$ and different values of coefficient $\beta$, $0.1$ and $0.5$, respectively. These intensity profiles are similar to dark solitons (albeit asymmetric in nature) and shows relative compression of pulses with the modulation of spectral filtering term $`\beta$'. Fig. \ref{fig2}(c) shows the profiles of corresponding frequency chirp $\delta\omega$ across the pulse of Lambert W-kink soliton at $z=0$. One can observe that frequency chirp saturates to negative value as the retarded time approaches its asymptotic limit and the amplitude of chirp can be controlled for judicious choice of parameter $`\beta$'.

\subsection{Moving front solitons}
To exemplify the existence of moving front solitons as exact solutions of this model,
let us consider the differential equation
in $\rho$
    \be\label{eq22}\rho'=c\rho\(1-\frac{\rho^2}{2b}\),\ee
where $b,c$ are real parameters. The explicit moving front soliton is given by \cite{S.N.,rpal}
    \be\label{eq25}\rho(\xi)=\sqrt{b\(1+\mbox{tanh}(c\xi)\)},\ee
for $b>0$.
\begin{figure}[ht!]
\begin{center}
\includegraphics[angle=0,scale=.60]{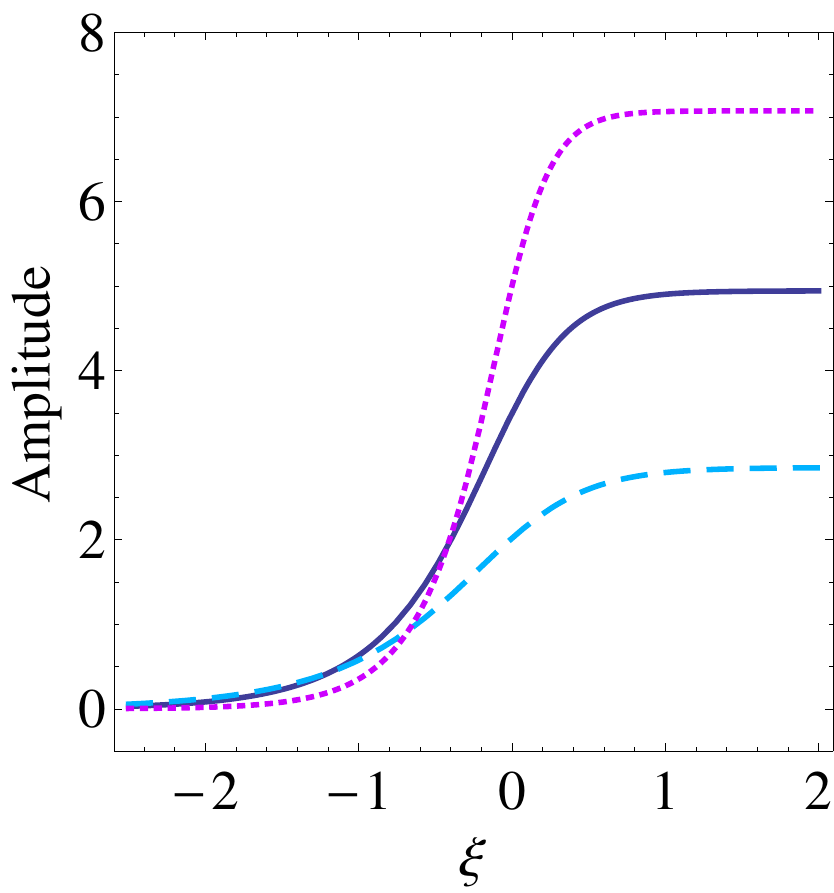}
\caption{\label{fig4} Amplitude profiles of kink solution for different values of $\beta$, $\beta = 0.3$ (thick line), $\beta = 0.4$
(dashed line) and $\beta = 0.5$ (dotted line). The other parameters used in the plots are $\epsilon =0.8,~\gamma =1,~\delta =-1$ and $T_r=0.2$.}
\end{center}
\end{figure}
The corresponding second-order differential equation for $\rho(\xi)$ reads
     \be\label{eq23}\rho''-\frac{3c^2}{4b^2}\rho^5+\frac{2c^2}{b}\rho^3-c^2\rho=0. \ee
Substituting Eq. (\ref{eq22}) into Eq. (\ref{eq13}) and comparing the resultant equation with Eq. (\ref{eq23}), the solution parameters found to be
    \be\label{eq24} c=\frac{-N\pm\sqrt{N^2-4Q}}{2},~~b=\frac{(N+4c)c}{2(R+Mc)},\ee
along with constraint on model coefficient $\nu=\frac{A^2}{2}+\frac{c M}{4b}-\frac{3c^2}{8b^2}$. Here, the parameters $M,N,Q$ and $R$ can be obtained from Eqs. (\ref{eq8})-(\ref{eq11}), using Eq. (\ref{eq15}), for different values of the model coefficients $\beta,~\epsilon,~\gamma,~\delta$ and $T_r$. For illustrative purpose, we choose the model coefficients $\epsilon>0$, $\gamma>0$ and $\delta<0$, just opposite to the case of Lambert W-kink solution, to depict the evolution of kink solutions. The amplitude profile of kink solution is shown in Fig. \ref{fig4} for $\epsilon =0.8,~\gamma =1,~\delta =-1,~T_r=0.2$ and different values of $\beta$.
\begin{figure}[ht!]
    \begin{center}
     \subfigure[]{
    \includegraphics[angle=0,scale=.5]{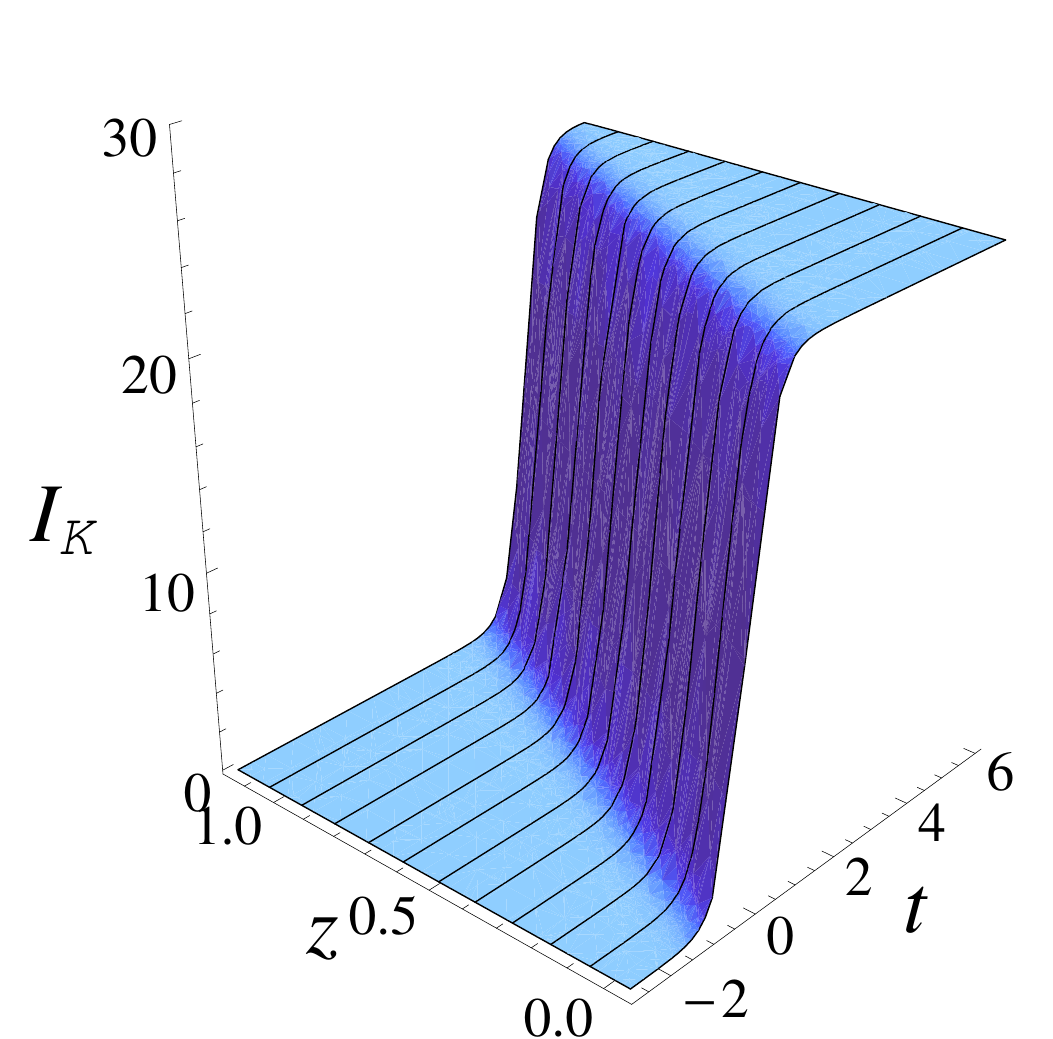}
    }
     \subfigure[]{
    \includegraphics[angle=0,scale=.5]{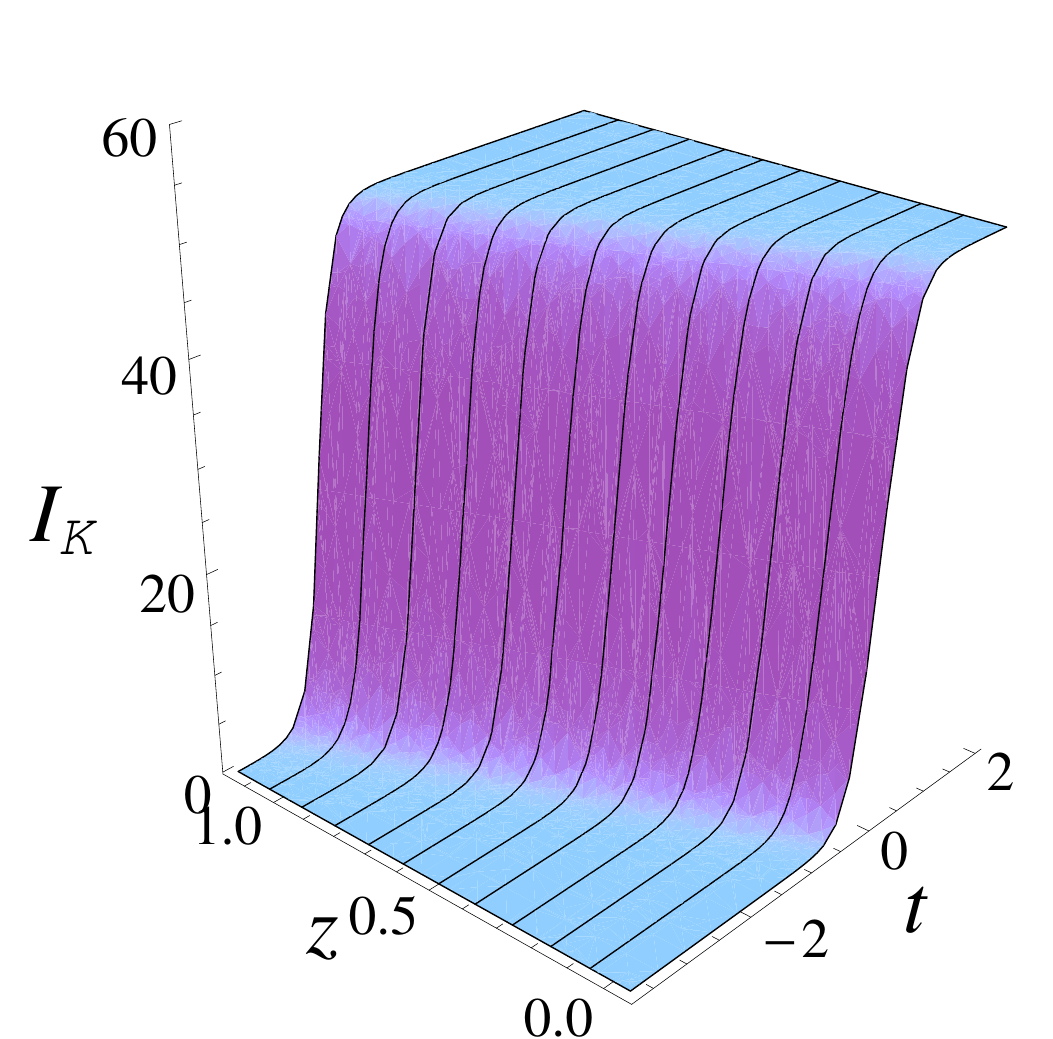}
    }
     \subfigure[]{
    \includegraphics[angle=0,scale=.7]{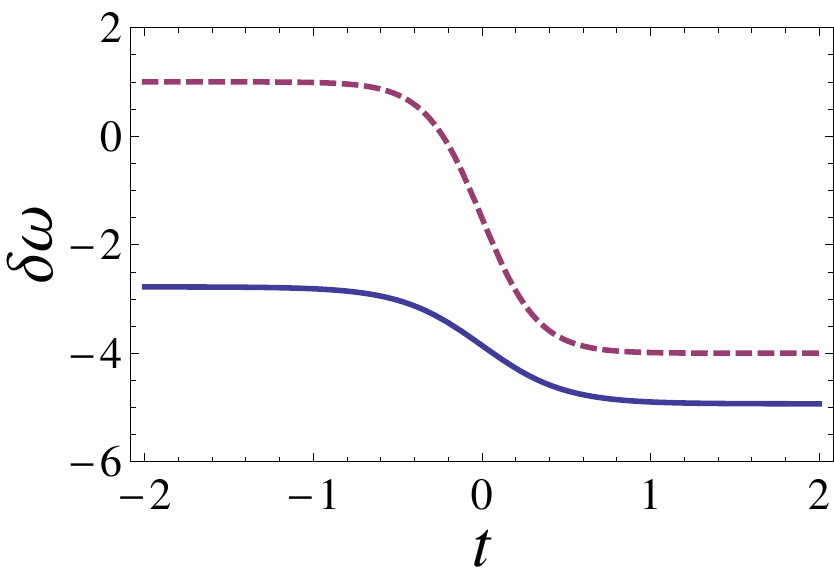}
    }
    \caption{\label{fig5} (a,b) Intensity profile of front soliton for different values of $\beta$, $0.3$ and $0.5$, respectively. (c) The corresponding chirp profiles for $\beta = 0.3$ (thick line) and $\beta = 0.5$ (dashed line). The other parameters are same as in Fig. \ref{fig4}.}
    \end{center}
\end{figure}
The expression of the intensity for these kink solution can be written as
    \be\label{eq28}I_K(z,t)=b\(1+\mbox{tanh}(c\xi)\).\ee
The corresponding chirping is given by
    \be\label{eq29}\delta\omega(z,t)=-\(\frac{u+2 \beta  T_r b\(1+\mbox{tanh}(c\xi)\)}{1+4 \beta^2}\). \ee
The intensity and chirp profiles for front solitons are shown in Fig. \ref{fig5} for
different values of $\beta$. For $\beta = 0.3$, the frequency chirp approaches negative value as $t\rightarrow\pm\infty$ while it approaches negative and positive value as $t\rightarrow~+\infty$ and $t\rightarrow~-\infty$, respectively, for $\beta = 0.5$.

\newpage

\section{Conclusion}
In conclusion, we have shown the existence  of exact explicit solution for the complex cubic-quintic Ginzburg-Landau equation in terms of Lambert W function or omega function, under the influence of  of intrapulse Raman scattering. Parameter domains are delineated in which these optical solitons exit in the ensuing model. It is observed that these optical solitons are possible for negative values of nonlinear gain and Kerr effect and positive value of constant gain. Whereas, no such restrictions are imposed on the moving fronts or optical shock-type solitons that are obtained as a byproduct of this model. We have observed that the intensity of these fronts has been doubled with a slight change in the value of the spectral filtering or gain parameter. The frequency chirp associated with these nonlinear waves has been identified. Furthermore, we have explicated the pivotal role played by this nonlinear chirp on the intensity of these waves. These results may be useful for experimental realization of undistorted transmission of optical waves in optical fibers and further understanding of their optical transmission properties. Finally, we hope that the exact nature of these nonlinear waves presented here may be profitably exploited in designing the optimal Raman fiber laser experiments.

\section{Acknowledgment}  A.G. gratefully acknowledges Science and Engineering
Research Board (SERB), Department of Science and Technology,
Government of India for the award of SERB Start-Up Research Grant
(Young Scientists) (Sanction No: YSS/2015/001803) during the course of this work. Nisha is thankful to SERB-DST,
India for the award of fellowship during the work tenure. We sincerely thank the referees for their useful comments.


\begin{thebibliography}{}

\bibitem{A.M.}A.M. Abourabia, R.A. Shahein, Eur. Phys. J. Plus 126 (2011) 23.
\bibitem{E.Kengne} E. Kengne, A. Lakhssassi, R. Vaillancourt, W.M. Liu, J. Math. Phys 53 (2012) 28.
\bibitem{A. Berti}A. Berti, V. Berti, Z. Angew. Math. Phys. 64 (2013) 1387.
\bibitem{R.J. Rivers}R.J. Rivers, J. Low Temp. Phys. 124 (2001) 41.
\bibitem{M. C.}M.C. Cross, P.C. Hohenberg, Rev. Mod. Phys. 65 (1993) 851.
\bibitem{M.S. Osman}M.S. Osman, Optik 156 (2018) 169.
\bibitem{Osman2}M.S. Osman, D. Lu, M.M.A. Khater, R.A.M. Attia, Optik 192 (2019) 162927.
\bibitem{akh}N. Akhmediev, A. Ankiewicz, Dissipative Solitons, Lecture Notes in Physics (Springer, Berlin, 2005).
\bibitem{Facao3}M. Fac\~{a}o, M.I. Carvalho, Phys. Rev. E 92 (2015) 022922.
\bibitem{facao4}M. Fac\~{a}o, M.I. Carvalho, Phys. Rev. E 96 (2017) 042220.
\bibitem{malomed}I.V. Barashenkov, S. Cross, B.A. Malomed, Phys. Rev. E. 68 (2003) 056605.
\bibitem{raju}T.S. Raju, K. Porsezian, J. Phys. A Math. Gen. 39 (2006) 1853.
\bibitem{amit}A. Goyal, Alka, T.S. Raju, C.N. Kumar, Appl. Math. Comput. 218 (2012) 11931.
\bibitem{osman01}J.G. Liu, M.S. Osman, A.M. Wazwaz, Optik 180 (2019) 917.
\bibitem{osman02}Y. Ding, M.S. Osman, A.M. Wazwaz, Optik 181 (2019) 503.
\bibitem{osman03}J.G. Liu et al., Appl. Phys. B 125 (2019) 175.
\bibitem{osman04}M.S. Osman, D. Lu, M.M.A. Khater, Results Phys. 13 (2019) 102157.
\bibitem{osman05}K.K. Ali, A.M. Wazwaz, M.S. Osman, Optik 2018 (2020) 164132.
\bibitem{4M. Mats} M. Matsumoto, H. Ikeda, T. Uda, A. Hasegawa, J. Lightwave Technol. 13 (1995) 658.
\bibitem{5H.A.}H.A. Haus, J.G. Fujimoto, E. Ippen, J. Opt. Soc. Am. B 8 (1991) 2068.
\bibitem{6J.D.}J.D. Moores, Opt.Commun. 96 (1993) 65.
\bibitem{kengne}E. Kengne, R. Vaillancourt, Can. J. Phys. 87 (2009) 1191.
\bibitem{saarlos}W.V. Saarlos, P.C. Hohenberg, Physica D 56 (1992) 303.
\bibitem{akh2}N. Akhmediev, A. Ankiewicz, Solitons: Nonlinear Pulses and Beams (Chapman $\&$ Hall, London, 1997).
\bibitem{crespo}J. Soto-Crespo, N. Akhmediev, Math. Comput. Simul. 69 (2005) 526.
\bibitem{J.Soto2}J. Soto-Crespo, N. Akhmediev, A. Ankiewicz, Phys. Rev. Lett. 85 (2000) 2937.
\bibitem{2001PRE}N. Akhmediev, J. Soto-Crespo, G. Town, Phys. Rev. E 63 (2001) 056602.
\bibitem{2001PLA}J. Soto-Crespo, N. Akhmediev, K. S. Chiang, Phys. Lett. A 291 (2001) 115.
\bibitem{2002PRL}S.T. Cundiff, J. Soto-Crespo, N. Akhmediev, Phys. Rev. Lett. 88 (2002) 073903.
\bibitem{tian}H.P. Tian, Z.H. Li, J.P. Tian, G.S. Zhou, J. Zi, Appl. Phys. B 78 (2004) 199.
\bibitem{latas}S.C. Latas, M.F. Ferreira, Opt. Lett. 35 (2010) 1771.
\bibitem{M.Facao1}M. Fac\~{a}o, M.I. Carvalho, S.C. Latas, M.F. Ferreira, Phys. Lett. A 374 (2010) 4844.
\bibitem{M.Facao2}M. Fac\~{a}o, M.I. Carvalho, Phys. Lett. A 375 (2011) 2327.
\bibitem{carvalho}M.I. Carvalho, M. Fac\~{a}o, Phys. Lett. A 376 (2012) 950.
\bibitem{latas2}S.C. Latas, M.F. Ferreira, M. M. Fac\~{a}o, Appl. Phys. B 116 (2014) 279.
\bibitem{2018}I.M. Uzunov, Z.D. Georgiev, T.N. Arabadzhiev, Phys. Rev. E 97 (2018) 052215.
\bibitem{2019}S.V. Gurevich, C. Schelte, J. Javaloyes, Phys. Rev. A 99 (2019) 061803(R).
\bibitem{anki}A. Ankiewicz, N. Akhmediev, Phys. Rev. E 58 (1998) 6723.
\bibitem{R.M.}R.M. Corless, G.H. Gonnet, D.E. Hare, D.J. Jeffrey, D.E. Knuth, Adv. Comput. Math. 5 (1996) 329.
\bibitem{S.R.}S.R. Valluri, R.M. Corless, D.J. Jeffrey, Can. J. Physics 78 (2000) 823.
\bibitem{Biswas2}A. Biswas, D. Milovic, R. Kohl, Inverse Problems Sci. Eng. 20 (2012) 227.
\bibitem{potential}A.M. Ishkhanyan, Phys. Lett. A 380 (2016) 640.
\bibitem{Azadeh}A. Amado, A. Mohammadi, arXiv:1906.08803v2 (2020).
\bibitem{chirp1}D. Grischkowsky, A.C. Balant, Appl. Phys. Lett. 41 (1982) 1.
\bibitem{chirp2}W. J. Tomlinson, R. H. Stolen, and C. V. Shank, J. Opt. Soc. Am. B 1 (1984) 139.
\bibitem{chirp3}G.P. Agrawal, M.J. Potasek, Opt. Lett. 11 (1986) 318
\bibitem{chirp4}M. Neuer, K.H. Spatschek, Z. Li, Phys. Rev. E 70 (2004) 056605.
\bibitem{chirp5}A. Blanco-Redondo, C. Husko, D. Eades, Y. Zhang, J. Li, T.F. Krauss, B.J. Eggleton, Nat. Commun. 5 (2014) 3160.
\bibitem{Alka}Alka, A. Goyal, R. Gupta, C.N. Kumar, T.S. Raju, Phys. Rev. A 84 (2011) 63830.
\bibitem{goyal}A. Goyal, V.K. Sharmaa, T.S. Raju, C.N. Kumar, J. Mod. Opt. 61 (2014) 315.
\bibitem{S.N.}S.N. Behra, A. Khare, Pramana-J. Phys. 15 (1980) 245.
\bibitem{rpal}R. Pal, A. Goyal, S. Loomba, T.S. Raju, C.N. Kumar, J. Nonlinear Opt. Phys. Mater. 25 (2016) 1650033.



\end{thebibliography}
\end{document}